# Balancing orbital effects and onsite Coulomb repulsion through Na modulations in Na$_x$VO$_2$

Xi Chen*, Hao Tang*, Yichao Wang, Xin Li[†]

John A. Paulson School of Engineering and Applied Sciences,

Harvard University, Cambridge, MA 02138

*: Equal contribution

[†]: Corresponding author: lixin@seas.harvard.edu

## Abstract

Vanadium oxides have been highly attractive for over half a century since the discovery of the metal insulator transition near room temperatures. Here Na$_x$VO$_2$ is studied through a systematic comparison with other layered sodium metal oxides with early 3d transition metals, first disclosing a unified evolution pattern of Na density waves through *in situ* XRD analysis. Combining *ab-initio* simulations and theoretical modelings, a sodium-modulated Peierls-like transition mechanism is then proposed for the bonding formation of metal ion dimers. More importantly, the unique trimer structure in Na$_x$VO$_2$ is shown to be very sensitive to the onsite Coulomb repulsion value, suggesting a delicate balance between strong electronic correlations and orbital effects that can be precisely modulated by both Na compositions and atomic stackings. This unveils a unique opportunity to design strongly correlated materials with tailored electronic transitions through electrochemical modulations and crystallographic designs, to elegantly balance various competition effects. We think the understanding will also help further elucidate complicated electronic behaviors in other vanadium oxide systems.

## Introduction

Vanadium oxide is one of the most widely studied prototypical strongly correlated materials. The discovery of the metal−insulator transition in VO$_2$ [1,2] triggers an extensive research regarding its origin and potential technological impact for over half a century. [3–14] Goodenough et. al [15] proposed a picture emphasizing the lattice and orbital effects, based on the single-particle theory of Peierls instability and the dimerization induced by the large crystal field splitting and hybridization gap. Later, Pouget, Zylbersztejn, and Mott emphasized the importance of electronic correlations, [16,17] to explain a new phase with alternated zigzag and dimerized chains by substituting only 0.2% of V in VO$_2$ by Cr. [18] Other theoretical models thus have been focused on either or both effects, [19–22] to try to understand the relationship between the two in the vanadium system.

Recently, experimental studies of the layered Na$_x$VO$_2$ (0 < x < 1) revealed important phenomena closely related to VO$_2$. [23] The layered Na$_{1/2}$VO$_2$ system was found to exhibit a magnetic phase transition concurrent with an insulator to metal transition at 322K. Peculiar V-trimers were also found to form at certain Na compositions, and *in-situ* XRD measurements identified a Na composition region with a continuous shift of superstructure peaks. NaVO$_2$ in fact belongs to a general material category of layered sodium transition metal oxides, i.e., NaTMO$_2$ (TM: transition metal), where all the 3d TM types have been synthesized and electrochemically tested for changing Na compositions, partly driven by the strong interest in the field of sodium ion batteries. However, this freedom in the choice of transition metal elements, where 3d electrons per TM ion can be adjusted, together with the freedom in the electrochemical control of Na compositions through de-intercalation, and the abundant interlayer stacking types and Na orderings, where the valence distribution and local environment of TM ions can be precisely tuned, also make the material system a unique platform to fine-tune the electron correlations and orbital interactions with unprecedented precision and flexibility that other metal oxide systems cannot match.

In this paper, we study Na$_x$VO$_2$ through the systematic comparison with other Na$_x$TMO$_2$ (TM for early 3d transition metal of Ti, V, Cr). Specifically, we note that similar to Na$_x$VO$_2$, a continuous shift of XRD peaks with Na compositions was also observed in Na$_x$TiO$_2$ [24] and our own Na$_x$CrO$_2$ experiments here. We will first solve the Na ordering patterns from the *in situ* XRD, giving a unified evolution pattern in the form of Na density waves (NDW). We then present a new picture that reconciles strong electronic correlations and orbital effects in Na$_x$TMO$_2$ using combined approaches of *ab-initio* DFT simulations and theoretical modeling. We propose a sodium-modulated Peierls-like transition mechanism for the bonding between TM dimers. More importantly, the unique trimer structure in Na$_x$VO$_2$ is shown to be a result of competition between the on-site Coulomb repulsion and orbital effects, which thus can be precisely modulated by a delicate collaboration of Na compositions and 3-dimensional atomic stackings to switch on and off, unveiling a unique opportunity to design strongly correlated materials with tailored electronic transitions through electrochemical modulations and crystallographic designs, to elegantly balance various competition effects.

## Methods

**DFT.** The geometric optimization and electronic structures are calculated by the density functional theory (DFT) and projector-augmented wave pseudopotential implemented in the Vienna ab initio simulation package (VASP) code with an energy cutoff of 520 eV. [31,32] For the normal accuracy calculation, we use the generalized gradient approximation (GGA) of the Perdew-Burke-Ernzerhof (PBE) functional. [33]

The on-site Coulomb repulsion is considered through GGA+U methods with the default U values of 0, 3.1, and 3.5 eV for Ti, V, and Cr, respectively, as reported by previous papers [34]. The energy converges to $10^{-5}$ eV for electronic iteration and the residue forces converge to 0.02 eV/A for ionic relaxation. The Monkhorst package for *k*-point mesh is used with *k*-points separation of fewer than 0.04 $A^{-1}$. [35] The DFT-D3 methods are used for van der Waals correction. [36,37] Results of GGA with different U values, GW methods [38,39] are compared for the $Na_xVO_2$ system to discuss the competition between bonding and onsite Coulomb repulsion. For GW calculation, we first obtain a one-electron basis set from a standard DFT calculation and use the basis set for the self-consistent QPGW algorithm [39] as the second step, where both the eigenvalues and one-electron orbitals are updated. The charge differential densities are calculated by a three-step procedure. (1) Calculate the charge density $n_1(r)$ distribution of the structure with the NDW and dimers/trimers. (2) Fix certain TM atoms to the lattice sites of dimer/trimer, and calculate the corresponding charge density $n_2(r)$ without forming the electronic dimer/trimer bondings. (3) Ensure that the considered atom(s) for both cases overlap, and calculate the charge density difference $n_1(r) - n_2(r)$. This function reflects the charge transfer during the formation of dimer/trimer bonds.

**1-D and 2-D Tight-binding models.** The origin of Na-modulated electronic interaction is analyzed by a tight-binding effective Hamiltonian model

$$\widehat{H} = \sum_{i,\alpha} \epsilon_{i,\alpha} c_{i,\alpha}^+ c_{i,\alpha} + \sum_{i,j,\alpha,\beta} t_{i,j}^{\alpha,\beta} c_{i,\alpha}^+ c_{j,\beta} + U \sum_i n_{i\uparrow} n_{i\downarrow} + \frac{1}{2} m\omega^2 \sum_i x_i^2 + \frac{1}{2} \sum_{i,j} v_{ij} N_i N_j \quad (1)$$

where $c^{(+)}$ denotes creation/annihilation operator of electrons, $n_{i\uparrow/\downarrow}$ represents electrons in TM d orbitals, $N_i$ denotes Na occupation in the corresponding interstices and $x$ is the displacement of transition metal ions. The orbital energy $\epsilon = -(e_0 N_0 + e_1 N_1)$ and hoping integral $t = t_0 + t_N N_t + t_d \Delta r$ are controlled by Na occupation, which gives rise to the Na induced electronic effect. $N_0$ and $N_1$ are the numbers of sodium ions directly above a TM atom and at the direction that the bonding *d*-orbitals orient to, respectively. $N_t$ is the number of sodium ions that are adjacent to the O atom on the bonding route. $\Delta r$ is the distance between two TM ions after considering the displacement of TM ions. $e_0$, $e_1$, $t_N$, $t_d$, and $t_0$ are coefficients in the model and are set as constants as intrinsic properties of the systems. Their values are determined through the comparison with the orbital polarization and dimer bond length in the DFT calculations.

A 1D-chain model with $d_{xy}$ bonding orbitals and the nearest $\sigma$-interaction is used with second-order perturbation methods (regarding the deviation of Na distribution from the uniformity perturbation) to derive the Fermi level instability (See Supplementary Information for detailed derivation).

To simulate the dimer formation for specific materials, the non-perturbation calculation considering the discrete distribution of Na ions is implemented by solving the tight-binding model accurately. For each distribution of transition metal and sodium ions, we have a set of $\epsilon_i$ and $t_{i,i+1}$ coefficients and can solve the band structure by the corresponding electronic Hamiltonian $H_e$. The electron occupations $n$ are solved iteratively. Given a set of values for $n_l$ at step $l$, we use $n_l$ to obtain the value for on-site Coulomb repulsion energy and plug the value into Eq. (1) as a constant. We then solve for the band structure and obtain the new occupations $n_{l+1}$. by diagonalizing the tight-binding part of Eq. (1) including the first three terms. The value of the U term we plugged in will contribute to the diagonal of the matrix for the Hamiltonian (i.e., it will add to the orbital energy $\epsilon_i$). Summing the band energy over the Brilliuon zone gives the total electronic energy. Together with the harmonic potential and Coulomb repulsion, we have

$$E_{tot} = 2S_{u.c.} \sum_{n,occ} \int_{BZ} \frac{d^2\mathbf{k}}{(2\pi)^2} E_n(\mathbf{k}) + \frac{1}{2}m\omega^2 \sum_i r_i^2 + \frac{1}{2}\sum_{i,j} v_{ij} N_i N_j \qquad (2)$$

where $S_{u.c.}$ is the area of the unit cell. Ionic relaxation of TM ions is implemented by minimizing the above function numerically, and the electronic structures are calculated by the convergent atomic positions. The trend of displacement and fluctuation of orbital occupation for different materials is calculated through this procedure in Fig. 3e.

The Hamiltonian of our 2D-model is:

$$\hat{H}_{2D} = \sum_{i,\alpha}(\epsilon_0 + \Delta\epsilon_{i,\alpha}) c_{i,\alpha}^+ c_{i,\alpha} + \sum_{i,\alpha}\sum_{j=i\pm i_\alpha} (t_0 + \Delta t_{i,j}^\alpha) c_{i,\alpha}^+ c_{j,\alpha} + \frac{1}{2}m\omega^2\sum_i r_i^2 + \frac{1}{2}\sum_{i,j} v_{ij} N_i N_j \qquad (3)$$

Where $\mathbf{i} = (i_1, i_2)$ goes through all TM atoms, $\alpha$ goes through $d_{xy}$, $d_{yz}$, and $d_{xz}$ orbitals, and $\Delta\epsilon_{i,\alpha}$ and $\Delta t_{i,j}^\alpha$ is determined by the Na environment as described in Eq. (1). It considers $t_{2g}$ orbitals (the first term) and $\sigma$ type hopping between them that points directly to each other (the second term). The harmonic potential of TM atoms and Na Coulomb repulsion terms are just to replace the 1D coordinates by 2D vectors compared to the 1D model. Similar to the 1-D case, atomic relaxation processes are used to consider the Na effect on the trimer formation, as briefly described by minimizing the total energy as below:

$$\min_{\{r_i\}} E_{tot} = \min_{\{r_i\}} \left[ 2S_{u.c.} \sum_{n,occ} \int_{BZ} \frac{d^2\mathbf{k}}{(2\pi)^2} E_n(\{r_i\},\mathbf{k}) + \frac{1}{2}m\omega^2\sum_i r_i^2 + \frac{1}{2}\sum_{i,j} v_{ij} N_i N_j \right] \qquad (4)$$

Where $E_n(\{r_i\}, \mathbf{k})$ is the electronic bands under the atomic configuration $\{r_i\}$, $S_{u.c.}$ is the area of the unit cell. The first term (electronic energy) is integrated over the first Brillouin zone and summed over all occupied bands. Onsite Coulomb repulsion U was not applied in the 2D-model, due to the complexity of 2D quantum many-body problem and that the model is sufficient without U as we only use it to simulate and understand the effect from specific Na environment that leads to trimer bonds in P2-Na$_x$VO$_2$, which agrees well with the DFT and experimental results. The different Na staking in O3 and P2 type materials are analyzed to quantify their modulation on the electronic structures. The energy curve as a function of TM displacement is calculated to reveal the Na driving force to the dimer formation and the spontaneous symmetry broken process during the trimer formation.

**XRD.** The *in-situ* lab XRD for NaCrO$_2$ was taken on Bruker D8 X-ray diffractometer equipped with a Mo source from a homemade in situ electrochemical cell with Be window. NaCrO$_2$ sample was synthesized following previous literature. [40] The *in-situ* cell was charged galvanostatically at C/50 rate between 2.0 and 4.5 V on Solartron 1287 with each XRD pattern scanned from 6.5° to 30.5° 2θ angle range (equivalent to 14.1° to 69.7° on Cu source) for 1 h, corresponding to 2% Na composition resolution per XRD pattern.

# Results

## 1. Na density wave in $Na_xTMO_2$

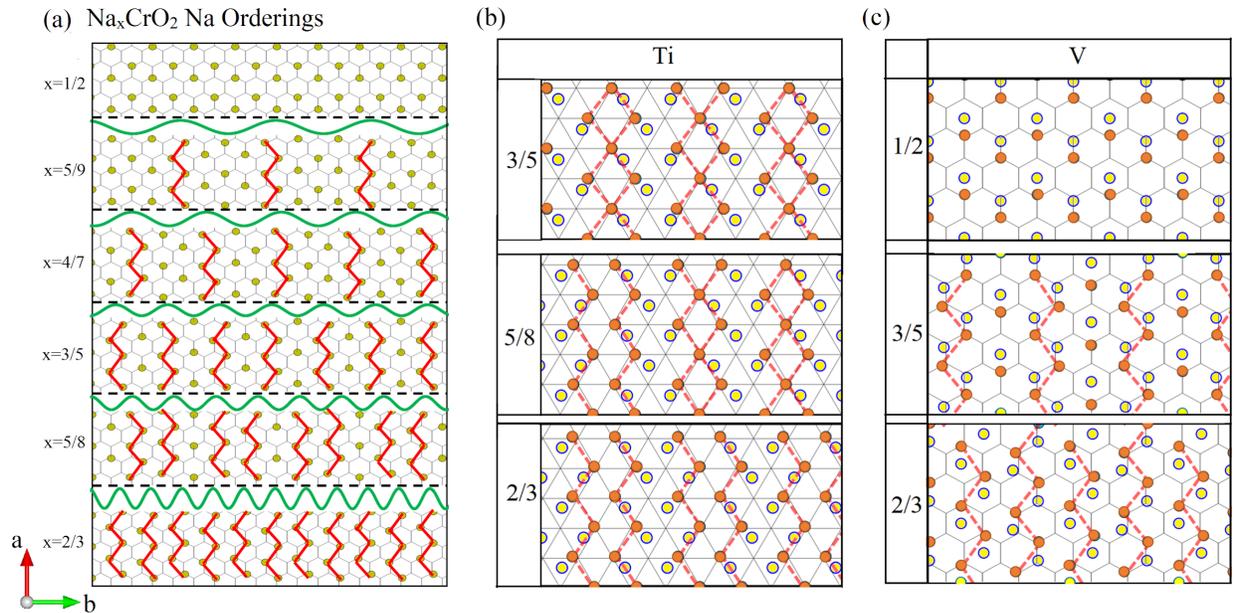

**Figure 1. Superstructure evolution and Na density wave (NDW) in $Na_xTMO_2$ (TM = Ti, V, Cr). (a)** Superstructure evolution in $Na_xCrO_2$ for $x$ in (1/2, 2/3). The Na ions are denoted by yellow disks on the hexagonal lattices. Antiphase boundaries and NDW with continuously varying wavelengths are denoted by red zigzag lines and green wave patterns, respectively. **(b)** Superstructure evolution of $Na_xTiO_2$. **(c)** Superstructure evolution of $Na_xVO_2$. The yellow and orange disks denote Na ions from two neighboring Na layers in the corresponding 2-layer supercells. The transition metal types and Na compositions are denoted on the top and left-hand sides of the figures, respectively. The stacking of the transition metal layers is along the $c$ direction in the O'3-Ti, P2-V, and P'3-Cr, where the notations of O'3, P2 and P'3 for various interlayer stackings are explained in **Table S1**.

Recent *in-situ* XRD experiments of $Na_xTMO_2$ (TM = Ti, V) clearly reveal the existence of incommensurate sodium orderings, with superstructure peak positions moving continuously and fastly as a function of sodium composition $x$. [23,24] This is analogous to the incommensurate electron or hole charge density waves found in other strongly correlated materials, including high temperature cuprate superconductors. [25] Here we first present a systematic comparison of the phase evolution in $Na_xTMO_2$ with early 3d TM = Ti, V, Cr. The Na incommensurate orderings of $Na_xCrO_2$ corresponding to the *in-situ* XRD patterns (**Fig. S1a-c**) are solved, with representative patterns at several discrete Na compositions shown in **Fig 1a**. At $x = 2/3$, which is the starting point of the superstructure peaks in electrochemical charge, Na ions form "zigzag" stripes along the $a$ direction, similar to the antiphase boundary Na ordering patterns predicted in $Na_xCoO_2$ recently [26]. With further charging to $x < 2/3$, more and more such zigzag stripes are replaced by another type of low-density stripes, until the low-density stripe region fully takes over at $x = 1/2$.

Here we view the Na ordering patterns in $1/2 < x < 2/3$ as Na density waves (NDW) with varying wavelengths. We find that similar NDW patterns (**Fig. 1b, c**) can also be identified from our analysis of previously reported *in situ* XRD results in $Na_xTMO_2$ with TM = Ti (**Fig. S1d**) [24] and V (**Fig. S1e**). [23] The fast and continuous evolution of the superstructure XRD peaks with changing Na compositions compared with the largely static background major *hkl* peaks can be understood as a two-component scattering, illustrated by orange arrows in **Fig. S1f,** with critical parameters listed in **Table S1**. The consistency of the XRD patterns between experiments and simulations based on our solved structures is illustrated in **Fig S1**. These structures with NDW are generally not the electrostatic ground states, which suggests that electronic effects overcome the Coulomb repulsion between sodium ions to enable the formation of the Na density wave.

## 2. Transition Metal Dimer Coupled with NDW

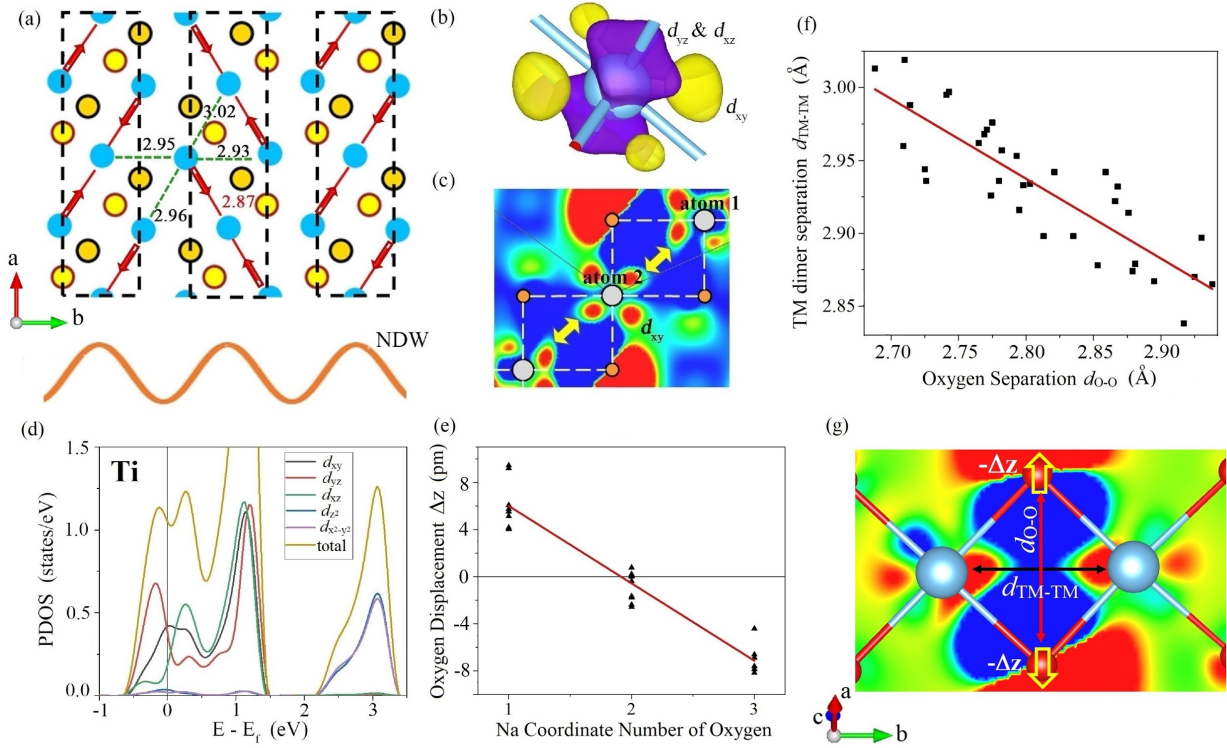

**Figure 2. Dimer and orbital interactions of NDW structures. (a)** Na density wave and the corresponding bonds & charge density wave of $Na_{2/3}TiO_2$. The Ti ions in a specific layer are represented by the blue disks, and the Na ions above and below the $TiO_2$ layer is represented by the dark yellow and bright yellow disks, respectively. The labeled distances between the Ti atoms are with the unit Å. **(b)** Positive and negative isosurface of charge density difference around Ti atoms. Electrons transfer from the purple $d_{yz}$ and $d_{xz}$ orbitals to the yellow $d_{xy}$ orbital. **(c)** Charge density difference between the NDW phase and the uniform phase without dimer (See Methods). Electrons are introduced to red areas which indicate $d_{xy}$ orbitals and σ bonds. **(d)** PDOS of $Na_{0.6}TiO_2$ projected on the $d$ orbitals of an atom involved in a dimer. The states from about -0.5 eV to 1.5 eV are from $t_{2g}$ orbitals, while the states from about 2 eV to 3.5 eV are from $e_g^*$ orbitals. **(e)** Displacement of oxygen atoms in the $z$-direction $vs$ the number of adjacent Na ions from statistics of a series of solved superstructures of $Na_xTiO_2$. Negative $\Delta z$ is the distortion toward the Na layer direction. **(f)** Relationship between the distance of the two oxygen atoms on different sides of the Ti ion layer and the corresponding TM dimer length. **(g)** Illustration of quantities defined in (e, f).

DFT relaxations of the $Na_xTMO_2$ structures with the solved Na orderings identify the formation of TM dimers, with shortened TM-TM bonds. The TM dimers are found to have varying strengths with Na compositions and TM types, and are closely related to the NDW pattern. **Figure 2a** shows the DFT relaxed structure of $Na_{2/3}TiO_2$. We observed the slight displacement of the TM atoms away from the octahedral center and the formation of the TM dimers at the Na density peak area of the NDW. With the formation of dimer bonds, electrons transfer from the $d_{xz}$ and $d_{yz}$ orbitals of both Ti ions in a dimer to their $d_{xy}$ orbitals **(Fig 2b)**. Since the $d_{xy}$ orbitals of the two bonding Ti ions point toward each other **(Fig 2c)**, it suggests that the dimers are bonded together in σ bonds.

The origin of the orbital interaction is further analyzed from the electronic structures. The *d*-bands of transition metals in $Na_xTMO_2$ are separated to the $t_{2g}$ and $e_g^*$ bands. The Ti ions in $Na_xTiO_2$ have occupation levels within $t_{2g}$ orbitals, and similar for other early 3d transition metals such as V and Cr. [27] The projected density of states (PDOS) of the $t_{2g}$ orbitals of the dimers, including the $d_{xy}$ and $d_{yz}$ orbitals for two different orientations, split into two peaks around the Fermi level as shown in **Fig 2d**, with the band below and above Fermi level as the bonding and antibonding states for TM dimers, respectively. The PDOS shows for the polarized *d* orbitals below the Fermi level, $d_{xy}$ shows higher occupation than $d_{xz}$ and $d_{yz}$, further suggesting the major contribution of the $d_{xy}$ orbital in the bonding state.

Importantly, we find that the dimer bonding formation is actually also modulated by local Na environments. The positively charged Na ions apply the negative local potential modulation for electrons in the $d_{xy}$ bonding orbitals, especially for those pointed directly to the Na ions (**Fig. S2a**), resulting in a higher local electron density (**Fig. S2b**). The modulation also pulls the negatively charged oxygen ions away from the Ti layer. These subtle crystal and electronic structural distortions further enhance the Ti-Ti bonding. The modulation effects are quantified based on the DFT simulations of such Na ordering supercells in **Fig. 2e** and **2f**. Our analysis shows the strong linear correlations between higher local Na ion density and larger oxygen displacement -$\Delta z$ away from the $TiO_2$ layer, and also the linear correlation between the separation of TM ions $d_{TM-TM}$, as a metric for the TM dimer strength, and that of oxygen ions $d_{O-O}$. These quantities are illustrated in **Fig. 2g.**

Furthermore, such interactions show unique material dependency. Although $Na_xCrO_2$ shows a similar electronic structure to $Na_xTiO_2$ (**Fig. S3a, 3c**), interestingly, O'3-$Na_{1/2}VO_2$ [28] and P2-$Na_{1/2}VO_2$ open bandgaps at the Fermi level (**Fig. S3b, 3d**), and display much stronger bonding interactions than the Ti and Cr systems.

## 3. Sodium-Modulated Peierls-like Transition

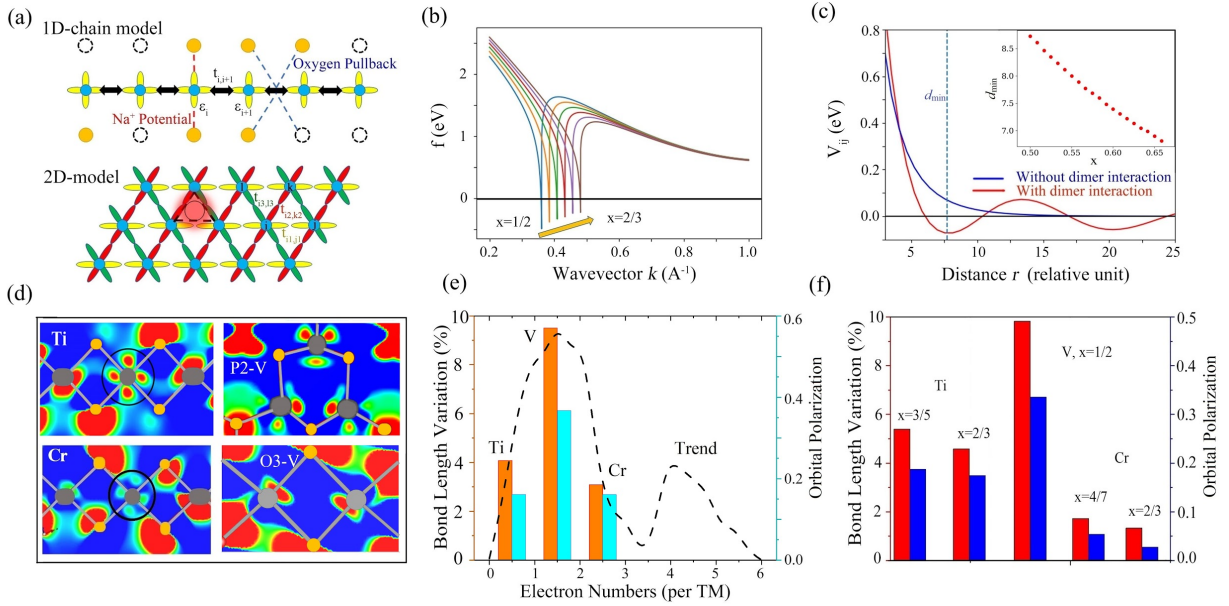

**Figure 3. Mechanism of the instability induced dimer and material trend**. **(a)** Illustration of the simplified 1D and 2D models. In the 1D model, a single $d_{xy}$ orbital (yellow lobes) is considered for each TM ion (green disks). The Na (yellow disks) distribution modulates both the illustrated orbital energy $\epsilon$ and the hopping integral $t$ of electrons. Empty circles indicate Na vacancies. In the 2D model, we consider the three $t_{2g}$ orbitals for each TM ion shown by yellow, green, and red lobes, respectively, as well as the hopping integrals between adjacent TM orbitals pointing to each other. **(b)** Energy coefficient $f_k$ of NDW formation in the wave vector space. **(c)** The effective potential of the interaction between Na ions at Na composition $x = ½$ given by the 1D model. The blue curve shows the screened Coulomb interaction (See Methods) between Na ions, while the red curve is given by our model with the electronic interaction added. Inset shows the continuous change of the Na-Na distance at the first minimum in the main plot. **(d)** Charge introduced into $d_{xy}$ bonding orbitals for O'3-$Na_{2/3}TiO_2$, O'3-$Na_{1/2}VO_2$, P2-$Na_{1/2}VO_2$, P'3-$Na_{1/2}CrO_2$. Electrons transfer to the more reddish area upon the formation of dimer or trimer bonds. **(e)** Tight binding model simulated dimer displacement or bond length variation (brown), and the dimer induced variation of electrons in bonding orbitals, i.e., orbital polarization (blue), for different materials. The dashed black line is the 1D-model calculated trend with the continuous change of electron numbers. **(f)** Comparison between bonding length change and bonding order of the dimer bond formation, i.e. orbital polarization, for different TM and Na composition x from DFT, where bonding length variation is measured by $(L - L_0)/L_0$, with $L$ and $L_0$ being the real bond length and the equilibrium one, respectively. The O'3-$Na_{1/2}VO_2$ is considered in (e) and (f).

Considering the quasi-1D nature of electronic localization discussed above, we start our modeling with the 1D-chain tight-binding model shown in **Fig 3a** to simulate the key mechanism of dimer formation by first considering the simplest case with the bonding interaction dominated by a single $d$-orbital. The model considers the following effective Hamiltonian, with the modulation of sodium environments on the orbital energy and tunneling coefficients between adjacent $d_{xy}$ orbitals as identified by DFT,

$$\widehat{H} = \sum_{i,\alpha} \epsilon_{i,\alpha}\, c^+_{i,\alpha} c_{i,\alpha} + \sum_{i,j,\alpha,\beta} t^{\alpha,\beta}_{i,j}\, c^+_{i,\alpha} c_{j,\beta} + U \sum_i n_{i\uparrow} n_{i\downarrow} + \frac{1}{2} m\omega^2 \sum_i x_i^2 + \frac{1}{2} \sum_{i,j} v_{ij} N_i N_j$$

where $c^{(+)}$, $n_{i\uparrow/\downarrow}$, $N_i \in \{0,1\}$, and $x$ represent the annihilation (creation) operator of electrons, electron number operator, Na occupation in the corresponding sites, and displacement of transition metal ions. The model includes orbital energy $\epsilon$, hopping integral $t$, *on-site* Coulomb repulsion $U$, harmonic potential for TM ions, and screened Na-Na repulsion $v_{ij}$. (See Methods and Supplementary Information for more details).

Our model first predicts NDW by describing it as an instability of the uniform Na distribution. By deriving the variation of total energy in response to the variation of Na density $n(k)$ in the momentum space, $\delta E[n] = 1/2 \sum_k f_k n(k)^2$, we find that negative coefficients $f_k$ emerge at a certain interval of $k$ (**Fig 3b**), indicating instability of the uniform Na distribution, because the formation of the NDW with the corresponding wave vector $k$ will decrease the energy of the system. Detailed derivation can be found in Methods. The most preferred NDW vector corresponds to the sharp valley in $f_k$ that moves continuously as a function of the Na composition $x$ (**Fig 3b**). The valley of potential occurs around 0.35-0.5 Å$^{-1}$, which roughly reflects the Na-Na distance that decreases from around 3 Å to 2 Å in real space as the Na composition increases from 1/2 to 2/3.

We then use our model to study the electronic and other ground-state properties by solving the tight-binding Hamiltonian at specific discrete Na distributions. The coupling between Na ions and d$_{xy}$ orbitals causes a Peierls-like transition, where the positions of TM ions in our model will be relaxed to form dimer bonds, and the effective interaction potential between Na ions at ground state shows several local minima (**Fig. 3c**) that continuously move with Na composition $x$ (**Fig. 3c inset**). Such modified potential drives the ground state Na ordering away from the electrostatic ground state, and the continuous evolution of local minima also corresponds to the continuous evolution of NDW upon (de)-intercalation of Na ions.

The strength of dimer bonds exhibits a trend for early 3d TM = Ti, V, Cr (**Fig 3d**). The intensity of dimer induced charge density is higher for Ti and V (red regions in the circled area) and lower for Cr, indicating that over-doping $t_{2g}$ orbitals will limit the bonding capability and hence weaken the strength of dimer bonds. Furthermore, both Na$_x$VO$_2$ phases show the uniqueness of V in orbital structures. The strong dimer distortion in O3-Na$_{1/2}$VO$_2$ makes the shape of d$_{xy}$ orbitals almost unrecognizable, while special V trimer bonds appear with a cloverleaf-pattern of bonding orbitals in P2-Na$_{1/2}$VO$_2$ (Red circled area in **Fig. 3d**).

The effect is further quantified by the orbital polarization or bonding order, which we define as the dimer induced variation of the electrons in bonding orbitals, and technically calculated as the number of bonding electrons minus that of antibonding ones. Our tight-binding model predicts the TM dependency of bonding strength measured by the bond length variation and orbital polarization for $Na_xTMO_2$ (**Fig. 3e**), which well matches the DFT simulations (**Fig. 3f**).

## 4. Uniqueness of $Na_xVO_2$ with $x$- and U-dependent Trimers

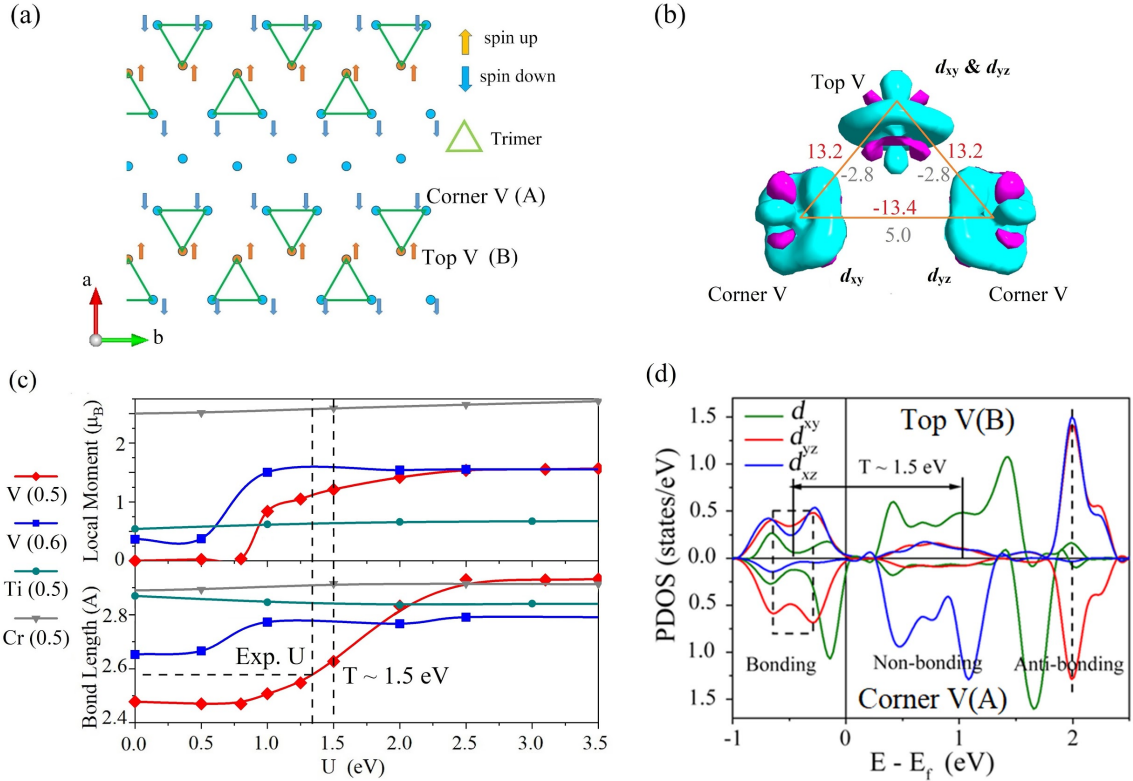

**Figure 4. Vanadium trimers of P2-$Na_{1/2}VO_2$.** (**a**) Optimized structure of P2-$Na_{1/2}VO_2$ by the GGA functional ($U = 0$). Trimer bonds of V atoms exist with opposite local magnetic moments at corners (A) and top (B) sites. (**b**) Inter-orbital charge transfer and magnetic interaction $J$ in trimer bonds. Charge transfers from the blue area to the purple area. The gray and red digits represent the value of $J$ in the unit of eV for low $U$ (1.5 eV) and high $U$ (2.5 eV), respectively. Positive and negative values indicate ferromagnetic (FM) and antiferromagnetic (AFM) couplings, respectively. (**c**) Average magnetic momentum on TM sites and dimer/trimer bond length and as a function of Hubbard $U$ value for TM = O'3-Ti, P2-V, and P'3-Cr at Na composition x = 1/2, as well as for P2-V at x = 0.6. (**d**) Projected DOS of P2-$Na_{1/2}VO_2$ calculated by the GW method, projected to the top V (B) and Corner V (A) in the trimer. The bands are divided into bonding orbitals, non-bonding orbitals, and antibonding orbitals in the trimer bonds.

The number of $d$ electrons, as the only variable in our model, shows a crucial role for the observed trend of the TM-TM coupling strength. The uniqueness for TM = V in **Fig 3e** and **3f** is also related to the filling levels of its $t_{2g}$ orbitals compared with other TM types. In **Fig. S3**, the Fermi level for Ti is at the left DOS peak and the dimer interaction strongly splits the peak, while $E_f$ stays at the right peak for Cr, where the dimer splitting effect is weak. In contrast, the $t_{2g}$ orbitals of TM = V are filled up to a valley in DOS, splitting out a small bandgap and making the $t_{2g}$ orbitals very susceptible to polarization into the bonding and antibonding orbitals with large separations. A typical band structure calculated from our 1D-model (**Fig. S4**) shows that the Fermi level of Ti falls in the lower-energy area of splitting, while the Cr falls to the higher-energy area, consistent with the DFT analysis.

The unique trimer bonds and bonding orbitals in P2-Na$_{1/2}$VO$_2$ (**Fig. 3d**) imply a special electronic origin with certain 2D nature, compared with the more 1D zigzag dimers. The DFT calculated P2-Na$_{1/2}$VO$_2$ structure is shown in **Fig 4a**, with 3/4 vanadium ions forming trimers. Atoms at the top and corner of the trimers are relaxed to opposite spin polarization directions in the ground state, forming ferromagnetic (FM) stripes along $b$ direction with the inter-stripe antiferromagnetic (AFM) coupling. Our trimer pattern is supported by previous XRD refinements of P2-Na$_{1/2}$VO$_2$, while our AFM ordering pattern here is consistent with the AFM type of ordering within VO$_2$ layers inferred from the magnetic susceptibility measurement. [23] Note that vanadium trimers are also observed in Li-excess Li$_x$VO$_2$. [29]

The electronic orbital interaction with the formation of trimer is visualized through the charge transfer shown in **Fig 4b**. Electrons transfer to the $t_{2g}$ orbitals involved in the trimer bonds from other $d$ orbitals during the bonding formation. The dominant bonding orbitals are the $d_{xy}$ and $d_{yz}$ ones of the corner V pointing to the top V, and the hybridization of the two corresponding $d$ orbitals of the top V atom (the purple ring-like orbital in **Fig. 4b**). The localized bonding interaction therefore mainly appears between the top and corner atoms according to the orbital ordering in the trimer.

More interestingly, the bonding and trimer formation is also strongly correlated with the change of magnetic coupling, and both are controlled by the on-site Coulomb repulsion $U$ in our DFT simulations. At $U = 0$, spin orderings other than the one shown in **Fig. 4a** are all unstable in DFT, and all are relaxed to the ordering in **Fig. 4a**. The trimer structure is maintained at $U = 1.5$ eV, where different spin orderings can be stabilized in DFT in order to compute the magnetic coupling $J$ based on the collinear model $E = \sum_{i,j} J_{ij}\sigma_i\sigma_j$. The solved magnitudes of $J$ (gray numbers in **Fig. 4b**) show the AFM coupling between the top and corner V atoms in the trimer, and the FM coupling between the two corner V atoms, which prefers the magnetic ordering in **Fig. 4a**. On the contrary, at $U = 2.5$ the trimers disappear, and all the $J$ coulings flip the sign. The FM coupling between the top and corner V ions in the trimer and AFM coupling between the two corner V ions lead to an FM ordering for all the V ions, while the two corner V ions are thus frustrated considering their AFM $J$ coupling.

A systematic study of sweeping the $U$ value reveals that the structural and orbital sensitivity to the $U$ value only happens in the P2-Na$_x$VO$_2$ system near x = 1/2, indicating a unique mixture of structural, orbital and strong electronic correlation effects. At x = 1/2, with $U$ value smaller than 0.8 eV, the strongest trimer bonds and the low net spin characterize the orbital-dominant condition with the electrons from the vanadium timers paired to form trimer bonds (**Fig. 4c**). With $U > 1$ eV, the trimer bond length (between the top and corner atoms) and local magnetic moment gradually increase, indicating the gradual broken of trimer bonds, until $U$ reaches approximately 2.5 eV. From the DOS calculation using *GW* method **(Fig. 4d),** which is in general considered to capture the strong electronic correlation more accurately, we identify that the average energy reduction of the bonding states is approximately 1.5 eV from the non-bonding states. Thus 1.5 eV becomes a critical value for the bonding energy of the trimer bonds to compete with the on-site Coulomb repulsion from the pair occupation in the bonding states, making the trimer bonds energetically sensitive to $U$ near the critical value.

By matching the trimer bond lengths from DFT and the experimental bond lengths at x = 1/2 [23] through sweeping $U$, we identify a value of $U_{exp} = 1.3$ eV that is slightly below the critical value of 1.5 eV. Therefore, it suggests that the experimental P2-Na$_{1/2}$VO$_2$ system slightly favors the bonding of trimer over on-site Coulomb repulsion. More importantly, the recipient of $U_{exp}$ to the critical $U$ value makes the trimer susceptible to external perturbations. It's also important to note that the critical $U$ value also decreases with increasing Na composition in P2-Na$_x$VO$_2$ at x > 0.5, as demonstrated at x = 0.6 (**Fig. 4c top**), suggesting the weakened bonding strength with Na doping that is consistent with the longer V-V bond length in the trimer (**Fig. 4c bottom**). At higher Na composition at $x = 0.67$, such trimer completely disappears at any $U$ values.

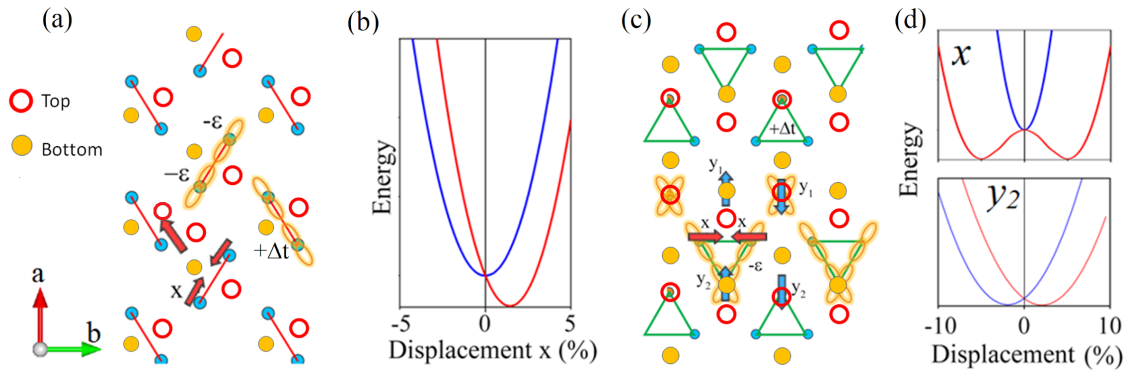

**Figure 5. Mechanism of trimer formation in P2-Na$_{1/2}$VO$_2$. (a)** Na-coupled TM orbital interaction in O'3-Na$_{1/2}$VO$_2$ with the ground state ionic dimers from our tight-binding models. **(b)** The local potential of the labeled ions in O'3-Na$_{1/2}$VO$_2$ along the local *x* axis in (a). **(c)** Ground state ionic trimers in P2-Na$_{1/2}$VO$_2$

from our tight-binding models. **(d)** The local potential of the labeled ions in P2-Na$_{1/2}$VO$_2$ along the local $x$ and $y$ axes in (c). Red and blue arrows in (a, c) indicate the displacement directions of several selected ions. Orange lobes illustrate the type of TM-TM bonding. $-\epsilon$ and $+\Delta t$ indicate the decrease of orbital energy and increase of the hopping integral induced by sodium. The results are consistent with our DFT simulations.

To further reveal the mechanism of the trimer bonding and the uniqueness of V trimers in P2-Na$_{1/2}$VO$_2$, we first compare P2-Na$_{1/2}$VO$_2$ with O'3-Na$_{1/2}$VO$_2$. The comparison of orbital effects in dimer and trimer bonds is shown in **Fig. 5**. The energy curve as a function of displacement $x$ and $y_2$ (all other degrees of freedom are optimized) shows that the O3 dimer bonds [30] are a Na-induced drifting force, while the P2 trimer bonds exhibit spontaneous breaking of inversion symmetry, leading to "dimers" in the crystal $b$ direction. The difference originates from the different interlayer stackings of Na **(Fig. 5ac)** as understood by our two-dimensional tight-binding model described in **Fig 3a**. In the O3 stacking system, the bonding strength of the dimer bonds denoted by red lines are enhanced by both the Na potential and the oxygen pullback effect, leading to lower orbital energy $\epsilon$ and stronger hoping integral and $t$, as illustrated in **Fig. 5a**. We can see that for every V dimer, the third V that forms a triangle with the dimer is pulled away from the dimer, causing no formation of the trimer. On the contrary, the Na effects in P2-Na$_{1/2}$VO$_2$ enhance the electronic bonding between the corner and top V atoms and lead to trimer bonds, as well as for Na composition $x$ at 0.6 that we checked in DFT simulations, where local Na environments close to P2-Na$_{1/2}$VO$_2$ still exist in the low-density area of the NDW in P2-Na$_{0.6}$VO$_2$.

In order for the trimer bonds to form among the three V ions, electrons from the top V ion need to overcome the on-site repulsion to occupy the three bonding orbitals. These orbitals are already half-filled by the electrons from the two corner V atoms with down-spins (**Fig. 4a**). With the formation of trimer, up-spins are injected to the bonding orbital, which matches the observation of the disappearing local magnetic moment at $U_{\text{exp}}$ (**Fig. 4c**) and the "singlet" state proposed for the V-trimer in Li excess Li$_x$VO$_2$ [29]. In contrast, the dimer bonds in these systems are insensitive to the $U$ value because no such extra pair occupation can be formed during the bonding process. Thus, the unique sensitivity to on-site Coulomb repulsion only exists for the trimer formation in P2-Na$_x$VO$_2$, where the competition between orbital and strong electronic correlation effects is balanced close to the critical value through Na modulations. More implicit modulation mechanisms, often applied through external parameters equivalent to the Na one here, are likely to play similar roles in modulating complicated electronic behaviors in other more controversial vanadium oxide systems.

# Conclusion

We studied the vanadium oxide system through a new perspective of Na modulations in a novel model system of layered Na$_x$VO$_2$. Through the systematic comparison with other Na$_x$TMO$_2$ (TM for early 3d transition metal of Ti, V, Cr), we observed similar continuous shift of XRD peaks with Na compositions, and we solved a unified evolution pattern in the form of Na density waves (NDW). We proposed a sodium-modulated Peierls-like transition mechanism for the bonding between TM dimers for all three TM species. Specifically for Na$_x$VO$_2$, we articulate quantitatively that the metal insulator transition in Na$_x$VO$_2$ is a result of delicate balances between competitive onsite Coulomb repulsion and orbital effects, which can be precisely modulated to switch on and off by alkaline ion compositions and 3-dimensional atomic stackings. Our findings unveil a new opportunity to design strongly correlated materials with tailored electronic transitions through electrochemical modulations, to elegantly balance various competition effects. We believe our understanding will also help further elucidate complicated electronic behaviors in other vanadium oxide systems.

# Acknowledgements

This work was supported by Dean's competitive fund of promising scholarship at FAS at Harvard University, and Harvard Data Science Initiative Competitive Research Award. Computations were supported by computational resources from the Extreme Science and Engineering Discovery Environment (XSEDE) and the Odyssey cluster supported by the FAS Division of Science, Research Computing Group at Harvard University. Use of the Advanced Photon Source at Argonne National Laboratory was supported by the U. S. Department of Energy, Office of Science, Office of Basic Energy Sciences, under Contract No. DE-AC02- 06CH11357.

# Supplementary Information for "Balancing orbital effects and onsite Coulomb repulsion through Na modulations in Na$_x$VO$_2$"

Xi Chen*, Hao Tang*, Yichao Wang, Xin Li†

John A. Paulson School of Engineering and Applied Sciences,

Harvard University, Cambridge, MA 02138

*: Equal contribution

†: Corresponding author: lixin@seas.harvard.edu

## 1. Detailed derivation of the tight-binding models

Here we show the detailed derivation of using a 1D-chain model with d$_{xy}$ bonding orbitals and the nearest $\sigma$-interaction with second-order perturbation methods regarding the deviation of Na distribution from the uniformity perturbation to derive the Fermi level instability. The distribution of sodium ions in the 1D-chain $N_i = N_0 + \delta N_i$ is the degree of freedom. $N_0$ represents the uniform distribution and the fluctuation $\delta N i$ can be understood as an average value of the grand canonical ensemble where the Na occupation at each site is either 0 or 1, with the average that can change continuously. Its Fourier transformation is as follow:

$$\delta N_i = \frac{1}{N} \sum_k \delta V_k e^{ikx_i} \tag{S1}$$

The instability of NDW can be studied through the energy functional to $\delta N_i$, which could be expanded to the second-order (the first order diminishes because the total number of sodium ions is a constant):

$$\delta E[\delta N] = \frac{1}{2}\sum_{ij} V_{ij}\delta N_i \delta N_j = \frac{1}{2N^2}\sum_{ij} V_{ij}(x_i - x_j)\sum_{k_1,k_2}\delta N_{k_1}\delta N_{k_2}e^{i(k_1 x_i + k_2 x_j)}$$

$$= \frac{1}{2N^2}\sum_{k_1,k_2}\delta N_{k_1}\delta N_{k_2}\sum_{i-j}e^{i(k_1-k_2)(x_i-x_j)/2}V_{ij}(x_i - x_j)\sum_{i+j}e^{i(k_1+k_2)(x_i+x_j)/2}$$

$$= \frac{1}{2N}\sum_{k_1,k_2}\delta N_{k_1}\delta N_{k_2}\sum_{i-j}e^{i(k_1-k_2)(x_i-x_j)/2}V_{ij}(x_i - x_j)\delta_{k_1,-k_2}$$

$$= \frac{1}{2N}\sum_{k}\delta N_k \delta N_{-k}\sum_{i-j}e^{ik(x_i-x_j)}V_{ij}(x_i - x_j)$$

$$= \frac{1}{2}\sum_{k}\left[\frac{1}{N}\sum_{l}e^{ikr_l}V(r_l)\right]|\delta N_k|^2 \tag{S2}$$

Define $f_k = \frac{1}{N}\sum_l e^{ikr_l}V(r_l)$, then we have the 2$^{nd}$ order expansion in the momentum space:

$$\delta E = \frac{1}{2}\sum_k f_k |\delta N_k|^2 \tag{S3}$$

To determine the coefficients $V_{ij}$ and $f_k$, we can calculate through the 2$^{nd}$ order perturbation theory regarding the deviation of Na density from the uniform $\delta N_i$ as the perturbation to derive the electronic energy:

$$\widehat{H}_e = \widehat{H}_{e0} + \delta \widehat{H}_e$$

$$= \left[\sum_i \epsilon_0 c_i^+ c_i + \sum_{i,j} t_0 c_i^+ c_j + U\sum_i n_{i\uparrow}n_{i\downarrow}\right]$$

$$+ \left[\sum_i \delta\epsilon_i c_i^+ c_i + \frac{1}{2}\sum_i \delta t_{i,i+1}(c_i^+ c_{i+1} + c_{i+1}^+ c_i)\right] \tag{S4}$$

Where the variation $\delta\epsilon_i$ and $\delta t_{i,i+1}$ is expanded for $N_i$ to the first order with coefficients $\partial\epsilon$ and $\partial t$, which are $\delta\epsilon_i = (\partial\epsilon)N_i$, $\delta t_{i,i+1} = \partial t$.

Parameters used in Fig. 3(b) and 3(c) are $t_0 = 2, \partial t = 0.5, \partial\epsilon = 0.7$, where $\partial\epsilon$ is the modulation to orbital energy when inducing 1 Na ion on either side. Coefficients are set as $e_0 = e_1 = 0.22, t_0 = 1, t_N = 0.15, m\omega/t_d = 5$ in Fig. 3(e). Parameters are set as $t_0 = 1.5, t_N = 0.5, t_d = 1$, $e_0 = 0.05, e_1 = 0.1, m\omega^2 = 10$ in Fig. 5. These values are determined through the comparison with the DFT calculations.

The 0$^{th}$ order condition corresponds to a totally uniform sodium environment, where $\epsilon_i$ and $t_{i,i+1}$ are constants:

$$E_0(k) = \epsilon_0 - t_0 \cos(kd), \qquad \psi_{0k}(k) = \frac{1}{\sqrt{N}} \sum_i e^{ikx_i} \phi_i(x) \tag{S5}$$

And then, we can calculate the 2$^{nd}$ order band energy

$$\delta E(k) = \langle \psi_k|\delta H_e|\psi_k\rangle + \sum_{k'} \frac{|\langle \psi_k|\widehat{\delta H_e}|\psi_{k'}\rangle|^2}{E_0(k) - E_0(k')}$$

$$= 0 + \frac{1}{N^2} \sum_{k'} \frac{\left|\sum_i e^{i(k'-k)x_i}[\partial\epsilon + (e^{ik'd} + e^{-ikd})\partial t]\, \delta N_i\right|^2}{E_0(k) - E_0(k')} \tag{S6}$$

The corresponding change of the total energy is therefore

$$\Delta E = \sum_{k,occ} \delta E(k) = \frac{1}{N^2} \sum_{k,occ} \left(\sum_{k',occ} + \sum_{k',vac}\right) \frac{\left|\sum_i e^{i(k'-k)x_i}[\partial\epsilon + (e^{ik'd} + e^{-ikd})\partial t]\, \delta N_i\right|^2}{E_0(k) - E_0(k')}$$

$$= 0 + \frac{1}{N^2} \sum_{k,occ}\sum_{k',vac} \frac{\left|\sum_i e^{i(k'-k)x_i}[\partial\epsilon + (e^{ik'd} + e^{-ikd})\partial t]\, \delta N_i\right|^2}{E_0(k) - E_0(k')}$$

$$= \frac{1}{2}\sum_{i,j}\left[\frac{2}{N^2}\sum_{k,occ}\sum_{k',vac} \frac{e^{i(k'-k)(x_i-x_j)}\left|\partial\epsilon + (e^{ik'd} + e^{-ikd})\partial t\right|^2}{\epsilon_0(k) - \epsilon_0(k')}\right]\delta N_i \delta N_j$$

$$= \frac{1}{N^2}\sum_{k,occ}\sum_{k',vac} \frac{\left|\partial\epsilon + (e^{ik'd} + e^{-ikd})\partial t\right|^2}{\epsilon_0(k) - \epsilon_0(k')}|N_{k'-k}|^2 \tag{S7}$$

The last two equations show the electronic part of $V_{ij}$ and $f_k$ through direct comparisons. Including the screened Coulomb interaction between sodium ions, we have:

$$V_{ij} = \frac{2}{N^2}\sum_{k,occ}\sum_{k',vac} \frac{e^{i(k'-k)(x_i-x_j)}\left|\partial\epsilon + (e^{ik'd} + e^{-ikd})\partial t\right|^2}{\epsilon_0(k) - \epsilon_0(k')} + v_{ij}$$

$$f_k = \frac{2}{N^2}\sum_{\substack{k'\in vac,\\ k-k'\in occ}} \frac{\left|\partial\epsilon + (e^{ik'd} + e^{-i(k'-k)d})\partial t\right|^2}{\epsilon_0(k'-k) - \epsilon_0(k')} + V_c(k) \tag{S8}$$

for coefficients in real space ($V_{ij}$) and reciprocal space ($f_k$) in Fig. 3c and Fig. 3b, respectively. The screened Coulomb potential is that $V_c(r) = e^{-r/\delta}/r$. The Coulomb potential $V_C(k)$ is its Fourier transformation cut-off at a large wave vector limit.

## 2. Supplementary figures

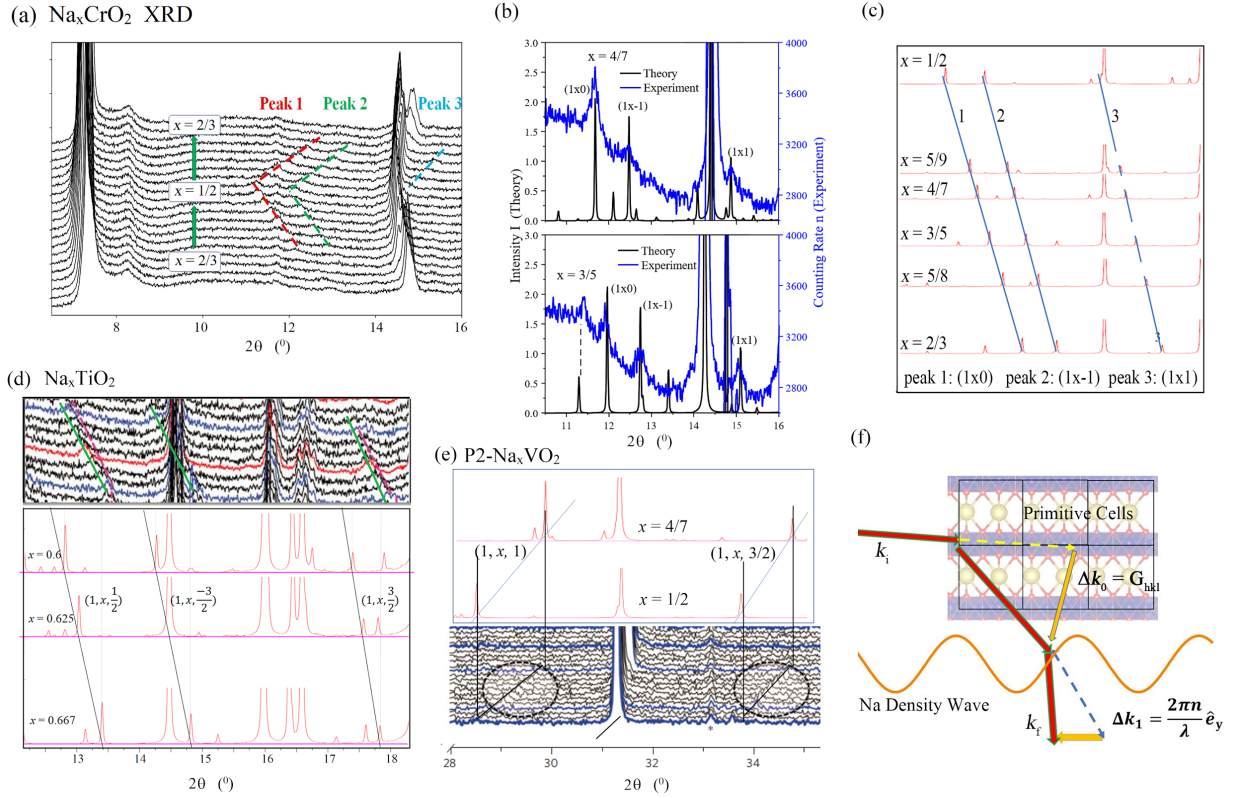

**Fig. S1.** (a) *In situ* XRD (Mo source) measurement of $Na_xCrO_2$ with three continuously-moving superstructure peaks. The charging process of the *in-situ* battery cell ends at $x = 1/2$. (b) Comparison of the *in situ* XRD (blue lines) and the spectra simulated from the solved superstructures (black lines) for two intermediate compositions $x = 4/7$ and $3/5$ of $Na_xCrO_2$. The major and superstructure peaks are all consistent, confirming the solved structures. (c) Simulated superstructure XRD peak evolution of $Na_xCrO_2$ from our solved structures. (d) Superstructure evolution of $Na_xTiO_2$ from both the *in situ* XRD (Mo source) experiments and solved structure simulations. The *in situ* XRD data is from the previous work [24] where three peaks are discovered. The solved Na orderings for $Na_xTiO_2$ are also fully supported by our *ex situ* synchrotron XRD analysis. (e) Superstructure evolution of $Na_xVO_2$ from both the *in situ* XRD (Cu source) experiments and solved structure simulations. The *in situ* XRD data is from the previous work. [23] (f) Illustration of the mechanism of how the XRD peak position continuously moves together with NDW wavelength during the (de)intercalation process.

|  | $Na_xTiO_2$ | $Na_xVO_2$ | $Na_xCrO_2$ |
|---|---|---|---|
| Lattice Type | O'3 | P2 | P'3 |
| Unit Cell | Monoclinic $a = 5.260, b = 3.037,$ $c = 5.697, \beta = 107.927°$ | Hexagonal $a = 2.865$ $c = 11.260$ | Monoclinic $a = 5.189, b = 2.981,$ $c = 5.868$ $\beta = 105.923°$ |
| Peak x Range | 0.58 ~ 0.66 | 0.50 ~ 0.57 | 0.50 ~ 0.66 |
| Superstructure Peak Index | 1: (1, x, 1/2) ~3.1 Å<br>2: (1, x, -3/2) ~2.8 Å<br>3: (1, x, 3/2) ~2.3 Å | 1: (1, x, 1) ~3.0 Å<br>2: (1, x, 3/2)<br>~2.6 Å | 1: (1, x, 0) ~3.5 Å<br>2: (1, x, -1) ~3.3 Å<br>3: (1, x, 1) ~2.7 Å |
| NDW $\lambda$ (Å) | $\dfrac{3.04}{x}$ | $\dfrac{2.87}{x}$ | $\dfrac{2.98}{x}$ |
| $\Delta k$ (Å$^{-1}$) | $2.07x$ | $2.19x$ | $2.11(2x - 1)$ |

**Table S1.** Comparison of properties of Na superstructures from the *in-situ* XRD measurements, including the type of Na sites, cell parameters, the composition range of moving superstructure peak in experiments, solved superstructure peak index as a function of x and the corresponding lattice plane distance, the NDW wavelength, and momentum change contributed by the scattering on the NDW. Note that the lattice type of O or P states that the $NaO_6$ local environment forms octahedral or prismatic geometry. O3, P3 or P2 classifies the interlayer stacking by every 3 or 2 repeating unit cells along *c*, while O'3, P'3 or P'2 labels the existence of monoclinic distortion. [41] The fast and continuous evolution of the superstructure XRD peaks with changing Na compositions compared with the largely static background major *hkl* peaks can be understood as a two-component scattering. $\Delta k_0$ on the underlying lattice with fixed momentum change of inverse lattice vector, and $\Delta k_1$ on the Na superstructure NDW with momentum change of $2\pi/\lambda \vec{e_y}$, where $\lambda$ is the wavelength of the NDW. From x = 2/3 to x = 1/2, $\lambda$ continuously increases, leading to a gradually smaller momentum change due to scattering on the NDW and the continuous shift of the XRD peaks toward the lower angle observed in experiments. Superstructure peaks are typically observed for *x* around 0.5 with the superstructure interplanar distance around 3 Å, whose (hkl) labels are shown in Table 1. The evolution patterns of all the peaks share a unified expression of Miller indices (1, *x*, *s*), where *s* is integer or half integer.

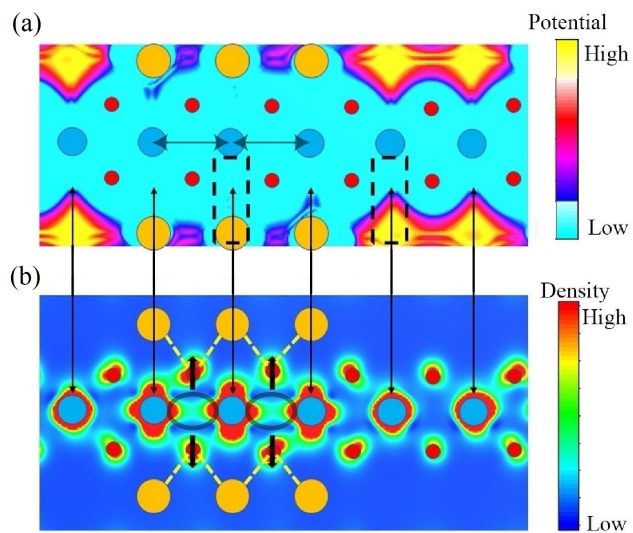

**Fig. S2.** Illustration of the modulation of Na ions on Ti-Ti bonding. (a) shows the local potential, where the Na points directly to Ti is boxed in dashed lines. (b) shows the probability density of $d_{xy}$ orbitals, where the black arrows illustrate the oxygen movement direction pulled away from the Ti ions by Na ions.

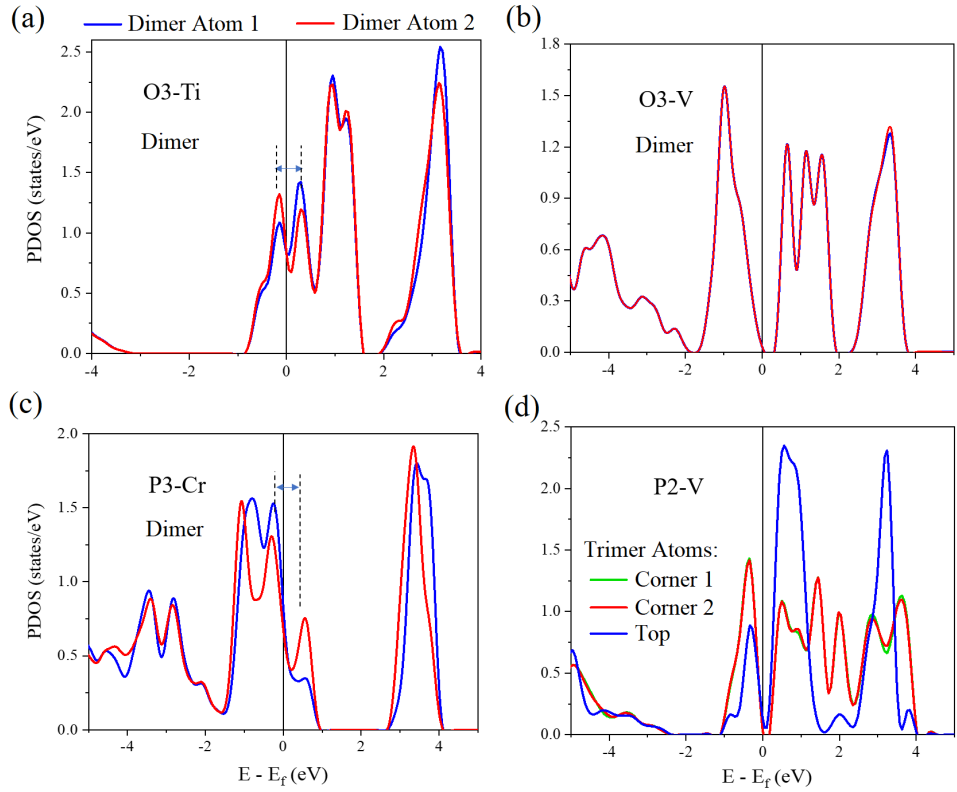

**Fig. S3.** Projected density of states (PDOS) of **(a)** O'3-$Na_{2/3}TiO_2$, **(b)** O'3-$Na_{1/2}VO_2$, **(c)** P'3-$Na_{1/2}CrO_2$, **(d)** P2-$Na_{1/2}VO_2$. The transition metal types and interlayer stackings are denoted in the figures. Most materials form dimers except P-2 $Na_xVO_2$ that forms the trimer. The calculated DOS are projected to TM ions in dimers and trimers as indicated by the legends. The splitting around the Fermi level is strong in Ti and V, but weak in Cr. The strong splitting in P2 and O3 $Na_xVO_2$ leads to the formation of small bandgaps.

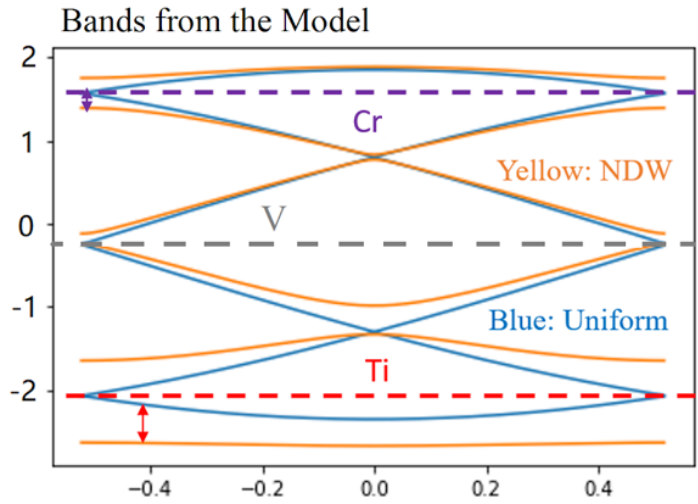

**Fig. S4** A typical band structure calculated from the 1D-model. The Fermi level of Ti falls in the lower-energy area of splitting, while the Cr falls to the higher-energy area.